\documentclass[aps,prd,twocolumn,nofootinbib,amsmath,amssymb,showpacs]{revtex4-1}

\usepackage{graphics}
\usepackage[usenames,dvipsnames]{color}
\usepackage{hyperref}

\usepackage[subnum]{cases}

\usepackage{lipsum}

\begin{document}
 
\title{Cold quark matter with heavy quarks and the stability of charm stars}

\author{Jos\'e C. {\sc Jim\'enez}}
\author{Eduardo S. {\sc Fraga}}

\affiliation{Instituto de F\'\i sica, Universidade Federal do Rio de Janeiro,
Caixa Postal 68528, 21941-972, Rio de Janeiro, RJ, Brazil}

\date{\today}


\begin{abstract}
We study the effects of heavy quarks on the equation of state for cold and dense quark matter obtained from perturbative QCD, yielding observables parametrized only by the renormalization scale. We investigate the thermodynamics of charm quark matter under the constraints of $\beta$ equilibrium and electric charge neutrality in a region of densities where perturbative QCD is, in principle, much more reliable. We also analyze the stability of charm stars, which might be realized as a new branch of ultradense hybrid compact stars, and find that such quark stars are unstable under radial oscillations.
\end{abstract}



\maketitle


\section{Introduction}
	
Heavy quark matter, i.e., quark matter including heavy flavors, could play a relevant role in extreme situations in the primordial quark-hadron transition \cite{Hajkarim:2019csy}. Experimentally, it is expected that the Facility for Antiproton and Ion Research (FAIR) with its Compressed Baryonic Matter (CBM) experiment will be able to produce charm quarks immediately after heavy-ion collisions with energies close to or above the charm threshold \cite{Ablyazimov:2017guv}. Cold quark matter also brings about the possibility of charm stars. Since the critical density required for their appearance is far above the limit imposed from causality together with the existence of two-solar mass neutron stars, as discussed in Refs. \cite{Lattimer:2010uk,Lattimer:2019eez}, such stars might be realized in nature only as a new branch of ultradense hybrid compact stars. 

As we go to higher values of quark mass, asymptotic freedom makes the perturbative quantum chromodynamics (pQCD) formalism more reliable \cite{Kapusta:2006pm}, so that this approach could be useful for heavy-ion collisions at low temperatures and high baryon chemical potentials as well as the physics of compact stars at ultrahigh densities, cases where charm quarks could play a role. One needs, then, to build the equation of state (EoS) for charm matter taking into account the constraints the system must respect below and above the charm threshold to generate matter configurations which are stable under electroweak interactions.

At high temperatures and zero quark chemical potentials, perturbative QCD was employed by Laine and Schr\"oder \cite{Laine:2006cp} to calculate the EoS including the charm quark contribution. These results were later compared to the ones provided by lattice QCD, including the charm and bottom contributions \cite{Borsanyi:2016ksw}, relevant for the study of the primordial Universe and its cosmological transitions \cite{Boeckel:2009ej,Boeckel:2011yj}. Considering nonzero chemical potentials for light and heavy quarks simultaneously implies some subtleties brought about by the heavy quarks at their mass thresholds. There, matching conditions should be imposed \cite{Rodrigo:1993hc}, having nontrivial effects on the possible values assumed by the renormalization scale $\bar{\Lambda}$.

In this work, we investigate cold quark matter with heavy quarks using in-medium pQCD. We build the framework for the case with two massive flavors, strange and charm quarks, and determine the equation of state. In practice, we extend the formalism developed for $N_{f}=N_{l}+1$ flavors in Ref. \cite{Kurkela:2009gj}, $N_{l}$ being the number of massless quarks and ``1" the massive flavor, to the case with any number of massive (heavy) flavors and explore their effects on the EoS of quark matter\footnote{Quark mass effects on the equation of state for cold quark matter were first considered  several decades ago \cite{Freedman:1977gz,Farhi:1984qu}, but only after two decades of being mostly ignored were computed within the modern $\overline{\rm MS}$ renormalization scheme \cite{Fraga:2004gz,Kurkela:2009gj,Fraga:2013qra,Kurkela:2014vha,Kurkela:2016was,Ghisoiu:2016swa,Gorda:2018gpy}.}. Given the equation of state, we discuss $\beta$ equilibrated and electrically neutral charm quark matter, and revisit the possibility of charm (quark) stars under the pQCD perspective. Charm stars were investigated in the past within the MIT bag model, being ruled out due to instabilities under radial pulsations \cite{Kettner:1994zs,Glendenning:2000,Haensel:2007yy} (see also Refs. \cite{SchaffnerBielich:1998ci,SchaffnerBielich:1997fx}). Using our results and the method of first-order coupled oscillation equations of Gondek \textit{et al.} \cite{Gondek:1997fd}, we also find that such quark stars are unstable under radial oscillations.

This work is organized as follows. In Sec. \ref{sec:framework}, we summarize the main aspects of the perturbative QCD formalism for $N_{f}=N_{l}+1$ flavors and presents our systematic extension to include heavy quarks in the framework. In Sec. \ref{sec:charm-matter}, we build the EoS for charm quark matter. In Sec. \ref{sec:charm-stars}, we solve the structure equations for charm stars and study their stability under radial acoustic perturbations. Section \ref{sec:conclusion} presents our summary and outlook.

\section{Framework}
	\label{sec:framework}

\subsection{Cold $N_{f}=N_{l}+1$ quark matter}

The equation of state for quark matter at high densities and zero temperature was first obtained in perturbative QCD by Freedman and McLerran \cite{Freedman:1976ub,Freedman:1977gz}; and Baluni \cite{Baluni:1977ms} in a modified momentum subtraction scheme over four decades ago (cf. also Refs.  \cite{Farhi:1984qu,Toimela:1984xy}). Later, it was computed in the modern $\overline{\rm MS}$ renormalization scheme for massless quarks in Refs. \cite{Blaizot:2000fc,Fraga:2001id,Fraga:2001xc,Andersen:2002jz,Vuorinen:2003fs}. 
These results were then extended to include the role of a massive quark\footnote{Some years ago, numerical interpolation techniques were used to match massless and massive dense pQCD equations of state \cite{Gangopadhyay:2017qjc}.} at two loops by Fraga and Romatschke \cite{Fraga:2004gz} and three loops by Kurkela \textit{et al.} \cite{Kurkela:2009gj}.

The latter framework was designed to deal with $N_{f}=N_{l}+1$ quark flavors, i.e., $N_{l}$ massless quarks plus $1$ massive quark. Originally, the massive flavor was chosen to be the strange quark in order to study its influence on the stellar structure of quark stars. The perturbative QCD thermodynamic potential up to the next-to-next-to-leading-order (NNLO) in the strong coupling $\alpha_{s}$, including the massless contribution plus a massive term, together with the mixed vacuum-matter (VM) diagrams and the corresponding ring terms, can be written as \cite{Kurkela:2009gj}
	\begin{equation}
	\Omega=\Omega^{m=0}(\vec{\mu})+\Omega^{m}(\tilde{\mu},m)+\Omega^{x}_{{\rm VM}}(\vec{\mu}, m)+\Omega_{{\rm ring}}(\vec{\mu}, \tilde{\mu}, m),
	\label{eq:potential}
	\end{equation}
  	where $\tilde{\mu}$ corresponds to the massive quark chemical potential, $\vec{\mu}~{\equiv}~(\mu_{1}, ...,\mu_{N_{l}})$ represents the vector chemical potential for the massless quarks, and $m$ is the physical (renormalized up to the same order in the strong coupling) mass\footnote{One must be careful when defining these quark masses since they only make sense in the UV regime, where asymptotic freedom takes place.} associated to the massive quark flavor. 	
	
Although not explicit in Eq. (\ref{eq:potential}), the thermodynamic potential depends also on a renormalization scale parameter $\bar{\Lambda}$, which is an additional scale generated by the perturbative expansion. Then, after solving the renormalization group equations in the $\overline{\rm MS}$ scheme for the strong coupling at this order, one obtains \cite{Fraga:2004gz}
     \begin{equation}
     \alpha_{s}(\bar{\Lambda})=\frac{4\pi}{\beta_{0}L}\left(
     1-\frac{2\beta_{1}}{\beta^{2}_{0}}\frac{\ln{L}}{L}\right),
     \label{eq:alphas}
     \end{equation}
where $\beta_{0}=11-2N_{f}/3$, $\beta_{1}=51-19N_{f}/3$, and $L=2\ln\left(\bar{\Lambda}/\Lambda_{\rm \overline{MS}}\right)$ with $\Lambda_{\rm \overline{MS}}$ being the $\overline{\rm MS}$ point (scale). Since $\alpha_{s}$ depends on the number of quarks and loops restricted by the active $N_{f}$, fixing the massive quark at some energy scale also depends on the number of flavors\footnote{For the massive quark running $m(\bar{\Lambda})$, see Ref. \cite{Vermaseren:1997fq}.}. 

One defines the renormalization scale parameter $\bar{\Lambda}$ in terms of the natural scale at ultrahigh densities, where quarks are massless, as being $\bar{\Lambda}=2\mu_{s}$ (with $\mu_{s}$ the strange quark chemical potential) for any $N_{f}$, and considers an uncertainty band defined by variations by a factor of $2$. Generically, it is convenient to write this parameter in the form $\bar{\Lambda}=X\sum_i{\mu_{i}}/N_{f}$, where the sum runs over all quark flavors that are present in the system, and the dimensionless parameter $X$ sits between $1$ and $4$ \cite{Kurkela:2009gj}.

	\subsection{Cold $N_{f}=N_{l}+N_m$ quark matter}

It is expected that at high densities not only the light quarks will be present in a system of quark matter, but also some heavy flavors. In QCD, heavy quarks are meant to be the ones satisfying $m\gg\Lambda_{\rm \overline{MS}}$, i.e., quarks with masses that are very large compared the QCD natural scale. Usually, their influence is neglected \cite{Kurkela:2009gj} in most calculations by invoking the $\textit{heavy-quark decoupling}$ theorem \cite{Symanzik:1973vg,Appelquist:1974tg,Bernreuther:1983zp}, which states that these quarks do not affect sizeably the observables calculated for light quarks, since their masses are of the order of the QCD energy scale.

We start by writing conveniently the total number of flavors in the form,
	\begin{equation}
	 N_{f}=\sum^{N_{m}}_{i=1}(N_{l}+1)^{(i)},
	 \end{equation} 
where $N_{m}$ is the number of massive quarks present in the system, and respecting the constraint $N_l + N_m=N_f$. So, we add at least one massless quark for each massive flavor included. For example, for charm quark matter, it will be convenient to write this sum over flavors as $N_{f}=(1+1)^{(1)}+(1+1)^{(2)}=2+2$, where $N^{(i)}_{l}=1$ and $N_{m}=2$. The usefulness of this way of writing $N_{f}$ will become clear after realizing the resemblance with the summing of massless and massive contributions to the total thermodynamic potential. Of course, this represents only a convenient way of writing the degrees of freedom at the level of the formalism. Additional physical conditions are needed in order to control when a heavy partner appears actively. Such conditions can be introduced by choosing appropriate values of the renormalization scale $\bar{\Lambda}$, depending on the chosen heavy flavor to be introduced in the system\footnote{Additional matching conditions on the renormalized QCD parameters should be imposed at the quark thresholds, i.e., on $\alpha_{s}(\bar{\Lambda}_{\rm thr})$ and $m(\bar{\Lambda}_{\rm thr})$, in order to account for their behavior at different values of $N_{f}$, depending on the energy scale of the problem \cite{Rodrigo:1993hc}.}. 
	
With these points in mind, we write the thermodynamic potential (up to NNLO) for $N_{l}$ massless and $N_{m}$ massive quarks as
	\begin{equation}
	\bar{\Omega}[{N_{f}}]=
	\sum^{N_{m}}_{i=1}\left\lbrace\Omega[{N^{(i)}_{l}}]
	+\Omega[{1^{(i)}}]\right\rbrace \; ,
	\label{eq:hQCDenergy}
	\end{equation}
where one must choose the number of massless flavors first when adding a massive one, so that
\begin{equation}
\Omega[{N^{(i)}_{l}}]~{\equiv}~(\Omega^{m=0}(\vec{\mu}))^{(i)} 
\end{equation}
is the massless contribution and 
	\begin{equation}
	\Omega[{1^{(i)}}]~{\equiv}~(\Omega^{m}+\Omega^{x}_{{\rm VM}}+\Omega_{{\rm ring}})^{(i)}
	\end{equation}
the mixed massive contribution, where $\vec{\mu}^{(i)}=(\mu_{1}, ...,\mu_{i})$ is the massless vector chemical potential, $\tilde{\mu}^{(i)}$ the massive (heavy) quark chemical potentials, and $m^{(i)}$ their corresponding masses. Here, $\Omega[...]$ indicates just the implicit parameter dependence (e.g. on $N_{f}$), whereas $\Omega(...)$ represents an explicit function dependence.
 
In the next section, we apply these results to the case of charm quark matter, and show that including heavy quarks makes the QCD thermodynamic potential less sensitive to the renormalization scale $\bar{\Lambda}$. In practice, its range of values is reduced in order to obtain a consistent thermodynamic transition between quark flavors, similar to results obtained in hot QCD \cite{Laine:2006cp,Graf:2015tda}.

	\section{Charm matter}
	\label{sec:charm-matter}	

In this section we consider the simplest case of heavy quark matter, charm quark matter, which is composed of light quark matter plus charm quarks. Of course, it can only be realized above a given critical charm chemical potential. As mentioned before, going to higher values of quark mass makes the perturbative QCD formalism more reliable. To build the EoS for charm matter we first need to establish the constraints this system must respect below and above the charm threshold, in order to generate matter configurations stable under electroweak interactions.
 
\subsection{Below the charm threshold: $N_{f}$=2+1}    
    
The condition of electric charge neutrality for a system with $N_{f}=2+1$ quarks (plus electrons) is given by
	\begin{equation}
	\frac{2}{3}n_{u}-\frac{1}{3}n_{d}-\frac{1}{3}n_{s}-n_{e}=0 \; ,
	\label{eq:sneutral}
	\end{equation}
where $n_{i}(\mu_{i})$ are the associated particle number densities for quarks and electrons in the system. The electron number density is well approximated, as usual, by that of a free Fermi gas, i.e., $n_{e}=\mu^{3}_{e}/(3\pi^{2})$.

Weak reactions among light quark flavors are given by
    \begin{equation}
    d~{\rightarrow}~u+e^{-}+\bar{\nu}_{e^{-}},\hspace{0.5cm}s~{\rightarrow}~u+e^{-}+\bar{\nu}_{e^{-}}
    \; ,
    \label{eq:sweak1}
    \end{equation}
    \begin{equation}
 {s+u}\leftrightarrow{d+u} \;.
 	\label{eq:sweak2}
    \end{equation}
and yield the following relations between chemical potentials:
	\begin{equation}
\mu_{d}=\mu_{s},\hspace{1cm}\mu_{u}=\mu_{s}-\mu_{e} \;.
	\label{eq:sweak3}
	\end{equation}
We neglect the neutrino chemical potential since their mean free path are large compared to the size of a typical compact star. 

Solving simultaneously Eqs. (\ref{eq:sneutral}) and (\ref{eq:sweak3}), one is able to write all the quark and electron chemical potentials in terms of only the strange chemical potential, $\mu_{s}$.
	
\subsection{Above the charm threshold: $N_{f}$=2+1+1} 
	
When $\mu_{s}$ crosses the charm quark threshold, the following weak equilibrium reaction is allowed to take place:
    \begin{equation}
   {u+d}\leftrightarrow{c+d} \;,
    \end{equation}
yielding the condition,\footnote{This is in contrast to the high temperature case of heavy ion collisions, where only thermal equilibrium of charm quarks is reached with the surrounding medium \cite{Torrieri:2010py}.}
	\begin{equation}
	\mu_{c}=\mu_{u} \;.
	\label{eq:cweak1}
	\end{equation}
The electric charge neutrality condition turns into
	\begin{equation}
	\frac{2}{3}n_{u}+\frac{2}{3}n_{c}-\frac{1}{3}n_{d}-\frac{1}{3}n_{s}-n_{e}-n_{\mu}=0 \;,
	\label{eq:cneutral}
	\end{equation}
where we have included free muons, with $n_{\mu}=(\mu^{2}_{\mu}-m^{2}_{\mu})^{3/2}/(3\pi^{2})$, which appear when $\mu_{\mu}>m_{\mu}=105.7~$MeV, where lepton number conservation allows us to write $\mu_{\mu}=\mu_{e}$. 

Again, by solving simultaneously Eqs. (\ref{eq:sweak3}), (\ref{eq:cweak1}), and (\ref{eq:cneutral}), we can express the quark and lepton chemical potentials only in terms of $\mu_{s}$. In the notation of Sec. \ref{sec:framework}, the charm matter thermodynamic potential corresponds to the case $N_{f}=(1+1)^{(1)}+(1+1)^{(2)}=(u+c)^{(1)}+(d+s)^{(2)}$ in Eq. (\ref{eq:hQCDenergy}).	

Now we need to fix the parameters entering the thermodynamic potential, i.e., running quark masses and strong coupling at some specific energy scale. Solving the renormalization group equations for the quark mass parameters up to second order in the strong coupling $\alpha_{s}$, one obtains the following results for the strange and charm quarks \cite{Vermaseren:1997fq}:
	\begin{eqnarray}
	\begin{aligned}
	m_{s}(\bar{\Lambda})=\hat{m}_{s}\left(\frac{\alpha_{s}}{\pi}\right)^{4/9}\hspace{4.3cm}\\
	\times\left(1+0.895062\left(\frac{\alpha_{s}}{\pi}\right) +1.37143\left(\frac{\alpha_{s}}{\pi}\right)^{2}\right) \;,
	\end{aligned}
	\label{eq:smass}
	\end{eqnarray}
	\begin{eqnarray}
	\begin{aligned}
	m_{c}(\bar{\Lambda})=\hat{m}_{c}\left(\frac{\alpha_{s}}{\pi}\right)^{12/25}\hspace{4.3cm}\\
	\times\left(1+1.01413\left(\frac{\alpha_{s}}{\pi}\right)+1.38921\left(\frac{\alpha_{s}}{\pi}\right)^{2}\right) \;,
	\end{aligned}
	\label{eq:cmass}
	\end{eqnarray}
with $\lbrace\hat{m}_{q}\rbrace$ being the renormalization group invariant quark masses, i.e., $\bar{\Lambda}$ independent\footnote{Expressing the quark masses in this way, one can see that their invariant masses can be fixed at independent energy scales, which is not obvious when using the quark mass function as in Ref. \cite{Kurkela:2009gj}.}.	
	
Since Eq. (\ref{eq:alphas}) for $\alpha_{s}$ tells us that different values of $N_{f}$ give different values of $\Lambda_{\overline{\rm MS}}$, by choosing  $\alpha_{s}(\bar{\Lambda}=1.5~{\rm GeV},~N_{f}=3,4)=0.336^{+0.012}_{-0.008}$ \cite{Bazavov:2014soa}, we obtain $\Lambda^{2+1}_{\overline{\rm MS}}=343^{+18}_{-12}~$MeV and $\Lambda^{2+1+1}_{\overline{\rm MS}}=290^{+18}_{-12}~$MeV, thus defining $\alpha^{2+1}_{s}(\bar{\Lambda})$ and $\alpha^{2+1+1}_{s}(\bar{\Lambda})$, respectively. Notice that the $\overline{\rm MS}$ renormalization scheme would require matching conditions for $\alpha_{s}$ as each quark threshold is crossed \cite{Rodrigo:1993hc}. However, since the corrections are very small, being a NNLO calculation, we have neglected them and only required flavor continuity of $\alpha_{s}$. Fixing the strange quark mass at $m_{s}(2~{\rm GeV}, N_{f}=3,4)=92.4(1.5)~$MeV \cite{Chakraborty:2014aca} give $\hat{m}^{2+1}_{s}~{\approx}~246.2~$MeV when using $\alpha^{2+1}_{s}$ in Eq. (\ref{eq:smass}), and $\hat{m}^{2+1+1}_{s}~{\approx}~243.7~$MeV with $\alpha^{2+1+1}_{s}$ also in Eq. (\ref{eq:smass}). Additionally, fixing the charm quark mass at $m_{c}(3{\rm GeV}, N_{f}=4)=0.9851(63)~{\rm GeV}{~\equiv~}m^{0}_{c}$ \cite{Chakraborty:2014aca}, gives $\hat{m}^{2+1+1}_{c}~{\approx}~3.0895~$GeV when using $\alpha^{2+1+1}_{s}$ in Eq. (\ref{eq:cmass}). We define $m^{0}_{c}$ as the vacuum charm mass for convenience later.

We assume that charm quarks are allowed in the system when
	\begin{equation}
	\mu_{c}=\mu_{s}-\mu_{e}>m^{\rm medium}_{c}>m^{0}_{c},
	\label{eq:charmcondition}
	\end{equation}
where $m^{\rm medium}_{c}$ is the (unknown) in-medium charm mass\footnote{An exact value for the in-medium charm mass at finite density is still not known, whereas its vacuum mass at some fixed energy scale, $m^{0}_{c}$, can be extracted from lattice calculations.}. 
Then, the renormalization scale parameter below and above the charm threshold is given by\footnote{Alternatively, one could choose independent values of $\bar{\Lambda}$ when going from $N_{f}=3$ to 4, the ``transition" point being found by a matching between strong couplings with different $N_{f}$ \cite{Rodrigo:1993hc,Graf:2016mzv}.}
	 \begin{numcases}
    {\bar{\Lambda}=}
      X\frac{(\mu_{u}+\mu_{d}+\mu_{s}+0)}{3} \;, &  $\mu_{s}~{\lesssim}~m^{0}_{c}$
      \label{A} \\
     X^{*}\frac{(\mu_{u}+\mu_{d}+\mu_{s}+\mu_{c})}{3} \;, &  
$\mu_{s}~{\gtrsim}~m^{0}_{c}$,
     \label{B}
  \end{numcases}
where the approximations in the inequalities of Eqs. (\ref{A}) and (\ref{B}) represent that fact that just before the threshold point the electron chemical potential takes its lowest value compared to the strange one, thus allowing us to make the approximation $\mu_{c}~{\approx}~\mu_{s}$. 

The only way to go from Eq. (\ref{A}) to Eq. (\ref{B}) continuously through the $N_{f}$ transition is by requiring the factors $X^{*}$ and $X$ to have the same range of possible values. To have them greater than $1$, one needs values greater than $4/3=1.3333...$ for both, which implies a reduction in the renormalization scale band of the EoS when heavy quarks are included, something already found in thermal perturbative QCD with charm quarks even at unusual low temperatures \cite{Laine:2006cp}.

\subsection{Thermodynamics for $N_{f}$=2+1+1} 
    
Using Eq. (\ref{eq:hQCDenergy}) with $N^{(1)}_{l}=1$ for the up, $N^{(2)}_{l}=1$ for the down, and $N_{m}=2$ for the strange and charm quarks, we have the following thermodynamic potential:
	 \begin{eqnarray*}
 \bar{\Omega}[N_{f}=2+1+1]=\left\lbrace\Omega[N^{(1)}_{l}=1]+\Omega[1^{(1)}]
 \right\rbrace+\\
 \left\lbrace\Omega[N^{(2)}_{l}=1]+\Omega[1^{(2)}]
 \right\rbrace \;,
	 \end{eqnarray*}
so that the flavors are counted as $N_{f}=(1+1)^{(1)}+(1+1)^{(2)}=(u+c)^{(1)}+(d+s)^{(2)}$. From this, one can identify the pressure as $P=-\tilde{\Omega}[N_{f}=2+1+1]$ and compute quark number densities using the standard thermodynamic relation\footnote{These derivatives are taken after fixing $X$ (see Ref. \cite{Haque:2014rua} for a discussion in the case of high temperature QCD).} $n_{f}=(dP/d\mu_{f})$. We define the total quark number density for charm matter, for a given $X$, as
    \begin{equation}
    n_{q}(\left\lbrace{\mu_{f}}\right\rbrace, X){\hspace{0.1cm}}{\equiv}{\hspace{0.1cm}}n_{u}+n_{d}+n_{s}+n_{c} \;,
    \end{equation}
and the total particle density as $n=n_{q}+n_{L}$, where $n_{L}=n_{e}+n_{\mu}$. 

    \begin{figure}[ht]
\resizebox*{!}{5.5cm}{\includegraphics{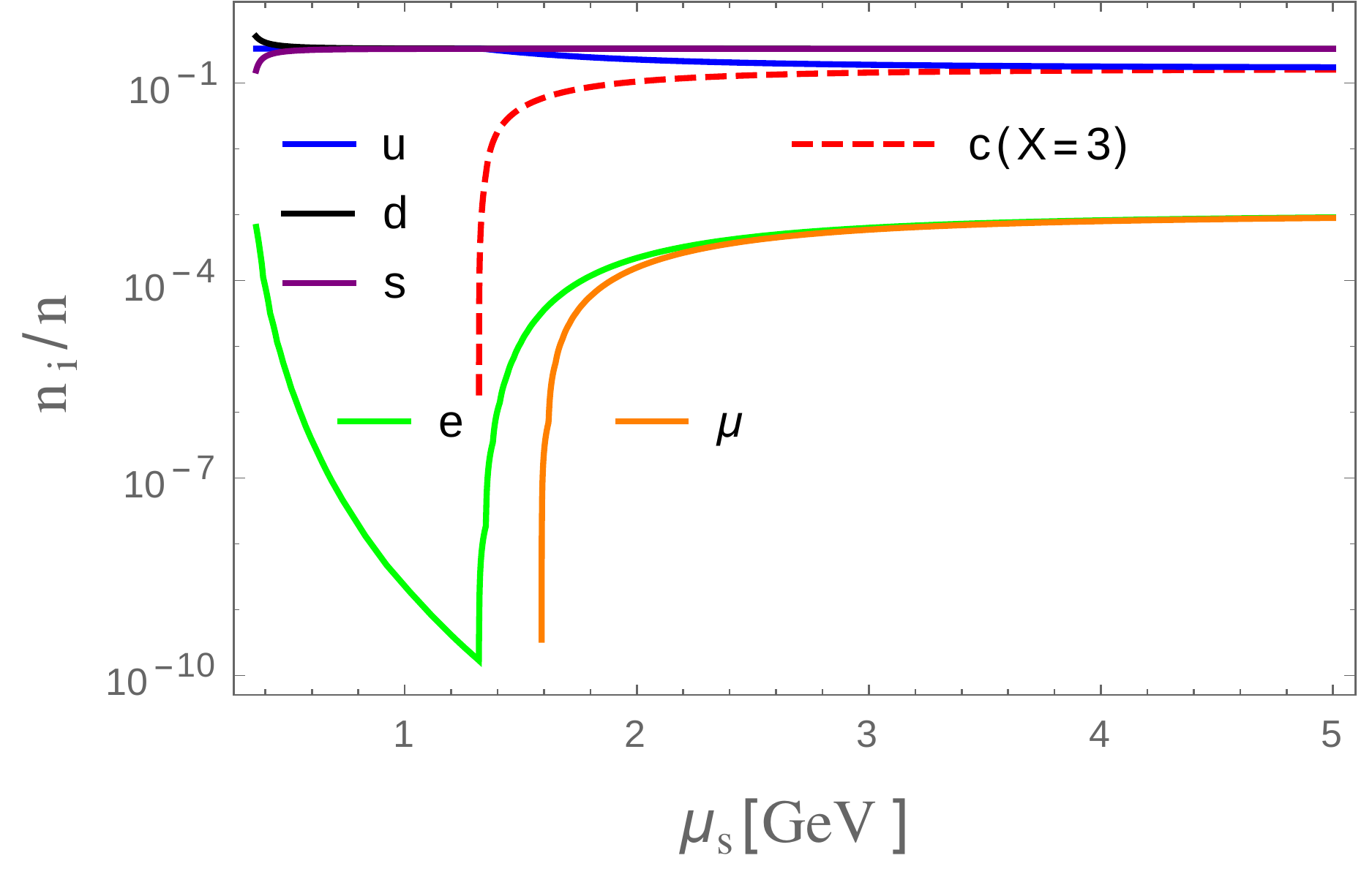}}
      \caption{\label{fig:par_densX3} Relative particle fractions for quarks and leptons, $n_{i}/n$, present in charm quark matter for $X=3$, where $i=u,d,s,c,e,\mu$. Above the charm quark threshold, the lepton fractions increase to ensure electric charge neutrality at high densities.}
    \end{figure}

In Fig. \ref{fig:par_densX3}, we show the behavior of the relative particle populations for our $\beta$ equilibrated and electrically neutral charm quark matter system in the case of $X=3$. Only above the charm threshold, charm quarks begin to contribute to the total number density, $n$. The location of the threshold depends on the value we choose for $X$ and is within $\mu_{s}~{\approx}~$1.2--1.4 GeV for the band we consider.
	
To build the total pressure, one should be careful with the fact that the derivatives of the thermodynamic potential give rise to terms which have the form of $\partial\alpha_{s}(\bar{\Lambda})/\partial\mu_{f}$ since $\bar{\Lambda}\propto{\mu}_{f}$, as can be seen in Eqs. (\ref{A})--(\ref{B}) (see also Ref. \cite{Kurkela:2009gj}). To keep thermodynamic consistency, one can take the quark and lepton number densities as the fundamental ingredients and build the other thermodynamic observables (e.g., pressure and energy density), imposing consistency on the number densities.
	
Thus, we define the total pressure of the system as
\begin{equation}
P(\mu_{s},X)=\sum_{f=u,c}P^{(1)}_{f}+\sum_{f=d,s}P^{(2)}_{f}
+\sum_{L=e,~\mu}P_{L} \;,
	\label{eq:pressure}
\end{equation}
where we have separated the contributions coming from $N_{f}=(u+c)^{(1)}+(d+s)^{(2)}$, defining each term as
\begin{eqnarray}
P^{(1)}_{f}(\mu_{s},X)=\int^{\mu_{s}}_{\mu_{0}(X)}d\bar{\mu}_{s}
 \left[ n_{f}\left(1-\frac{d\mu_{f}}{d\bar{\mu}_{s}}\right)\right] \;,
\end{eqnarray}
\begin{eqnarray}
\begin{aligned}
P^{(2)}_{f}(\mu_{s},X)=\int^{\mu_{s}}_{\mu_{0}(X)}d\bar{\mu}_{s}~n_{f} \; ,
\end{aligned}
\end{eqnarray}
and the lepton contribution as

\begin{eqnarray}
\begin{aligned}
P_{L}(\mu_{s},X)=\int^{\mu_{s}}_{\mu_{0}(X)}d\bar{\mu}_{s}
~n_{L}\frac{d\mu_{L}}{d\bar{\mu}_{s}}\; ,
\end{aligned}
\end{eqnarray}
including strange quarks even at zero pressure, from which we start the integration of the particle densities, and adding the charm and leptons when crossing their respective thresholds.   

We define the energy density as
\begin{equation}
\epsilon(\mu_{s},X)=-P+\sum_{f=u,c}\epsilon^{(1)}_{f}
+\sum_{f=d,s}\epsilon^{(2)}_{f}
+\sum_{L=e,\mu}\epsilon_{L} \;,
	\label{eq:energydensity}
\end{equation}
where the quark and lepton contributions are
\begin{eqnarray}
\epsilon^{(1)}_{f}(\mu_{s},X)=\left[ \mu_{s}-\mu_{e}(\mu_{s}) \right]n_{f}(\mu_{s}) \;,
\end{eqnarray}
\begin{eqnarray}
\begin{aligned}
\epsilon^{(2)}_{f}(\mu_{s},X)=\mu_{s}n_{f}(\mu_{s}) \;,
\end{aligned}
\end{eqnarray}
\begin{eqnarray}
\begin{aligned}
\epsilon_{L}(\mu_{s},X)=\mu_{L}(\mu_{s})n_{L}(\mu_{s}) \;.
\end{aligned}
\end{eqnarray}
Following this recipe we can build (numerically) the EoS, $P=P(\epsilon)$, by combining Eqs. (\ref{eq:pressure}) and (\ref{eq:energydensity}) for a given $X$.

    \begin{figure}[ht]
      \begin{center}
	\resizebox*{!}{5.5cm}{\includegraphics{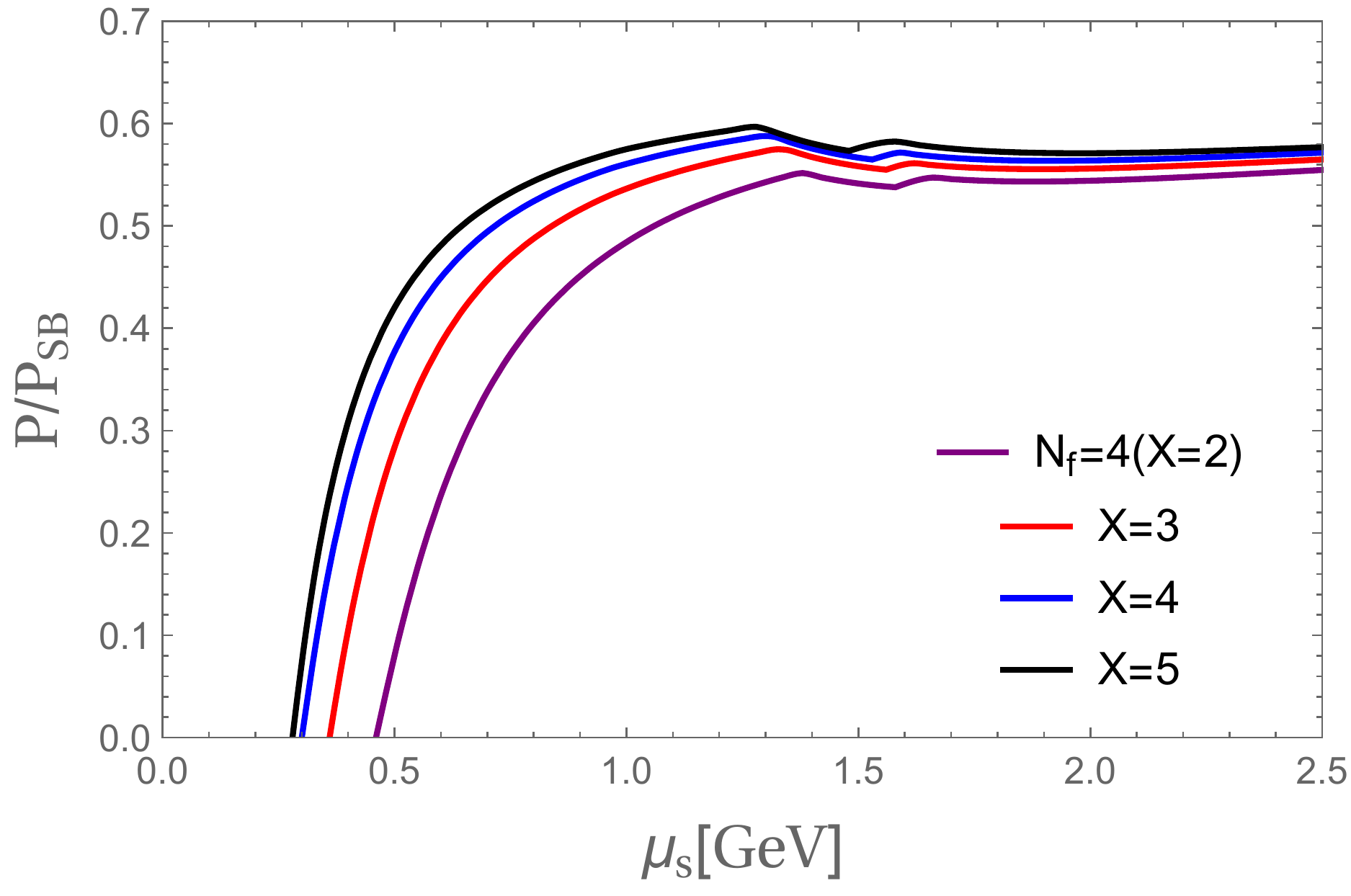}}
     \end{center}
      \caption{\label{fig:ChPressure_norm} Total pressure for a system of $N_{f}=2+1+1$ quarks plus leptons in $\beta$ equilibrium and electrically charge neutral normalized by the Stefan-Boltzmann massless free gas with $N_{f}=4$. We include the $X=5$ only to verify how the EoS depends on $X$ when including one additional massive flavor.}
    \end{figure}

In Fig. \ref{fig:ChPressure_norm}, we plot the total pressure for charm matter normalized by a Stefan-Boltzmann gas of quarks with $N_{f}=4$ as a function of the strange quark chemical potential. From this plot, it is easy to infer the contribution of each degree of freedom, at their thresholds, to the EoS for cold quark matter, and one can see the usual behavior of the pressure for $N_{f}=2+1$ at intermediate densities, followed by a kink representing the charm threshold which softens the total (normalized) pressure. The charm quark contribution reduces the renormalization-scale uncertainty band for $X$ at high densities, which also affects the behavior of the EoS a lower densities, a feature which would be difficult extract from the pressure-density plane. An additional kink appears due to the muons. So, the charm EoS is largely softened, generating an apparent instability which could have astrophysical effects. In particular, it suggests the possibility of another kind of ultradense compact star: charm stars.

\section{Are charm stars stable?}
\label{sec:charm-stars}

\begin{figure}[ht]
      \begin{center}
	\resizebox*{!}{5.5cm}{\includegraphics{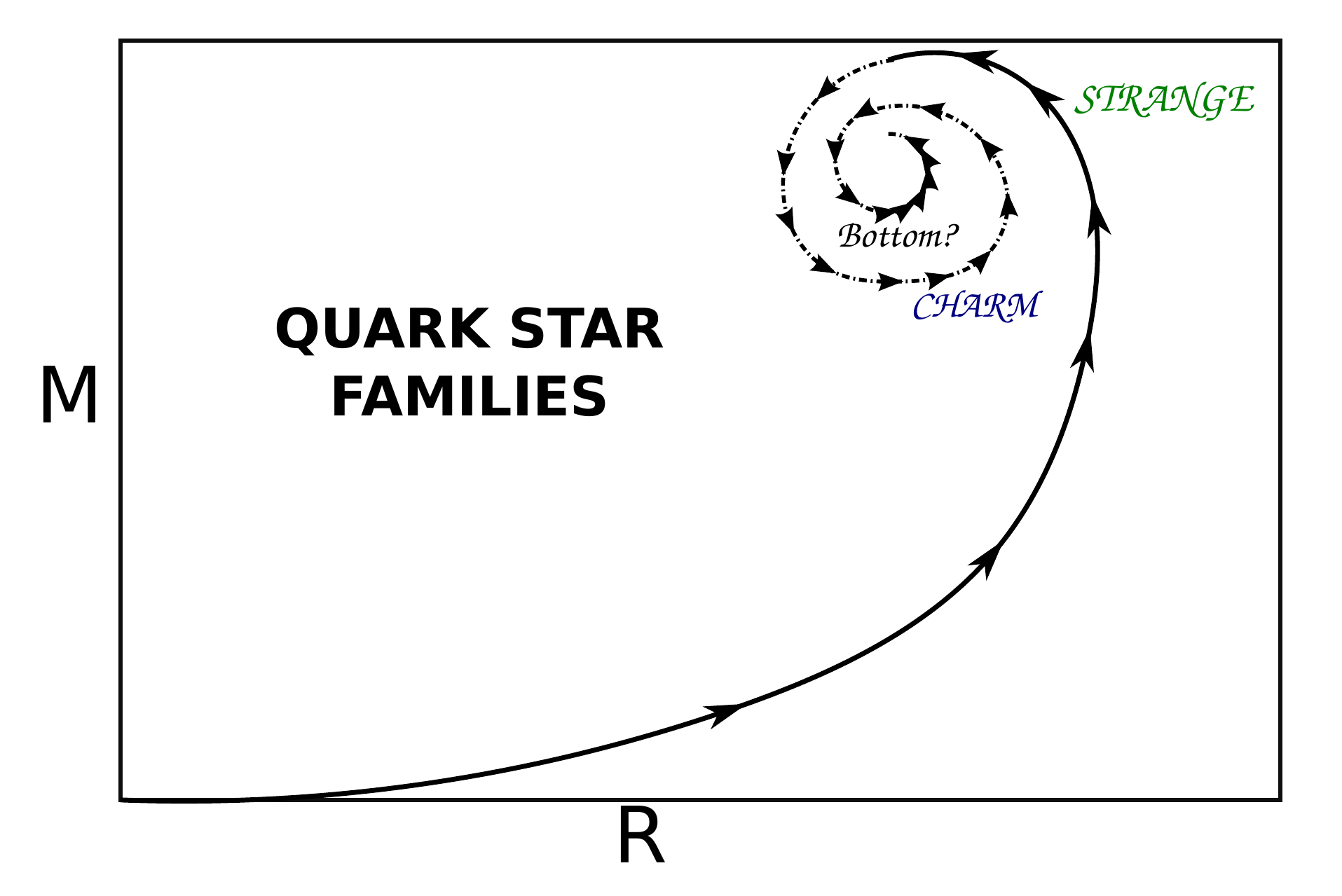}}
     \end{center}
      \caption{\label{fig:MRcharm} Cartoon of mass-radius diagram for quark star families in hydrostatic equilibrium. The only stable branch seems to be the strange (continuous-arrowed line), whereas ultrahigh density stars are usually considered unstable against radial pulsations (dashed-arrowed line).}
    \end{figure}

Although charm stars are excluded as two-solar mass neutron stars \cite{Lattimer:2010uk} given the high critical density required for their appearance, they might be present as a new branch of hybrid compact stars. The first quantitative study of the possibility of the existence of charm stars, i.e., strange stars satisfying the Bodmer-Witten hypothesis and having finite charm quark fractions at their cores, was carried out more than two decades ago; see Ref. \cite{Kettner:1994zs} (see also Ref. \cite{SchaffnerBielich:1998ci}). The star bulk was described using the simplest version of the MIT bag model. After performing the stability analysis, the conclusion was that charm stars would be unstable (see Fig. \ref{fig:MRcharm} for an illustration of standard quark stars).

We revisit this question using our first-principle perturbative QCD description for the EoS for charm quark matter and also restrict our analysis to the simple case with no hadronic mantle. So, we choose the parameter space to be in the range $X\geq3$, which satisfies the Bodmer-Witten hypothesis, as shown in Ref.\cite{Kurkela:2009gj}, and perform the stability analysis as follows.

    \begin{figure}[ht]
      \begin{center}
	\resizebox*{!}{5.5cm}{\includegraphics{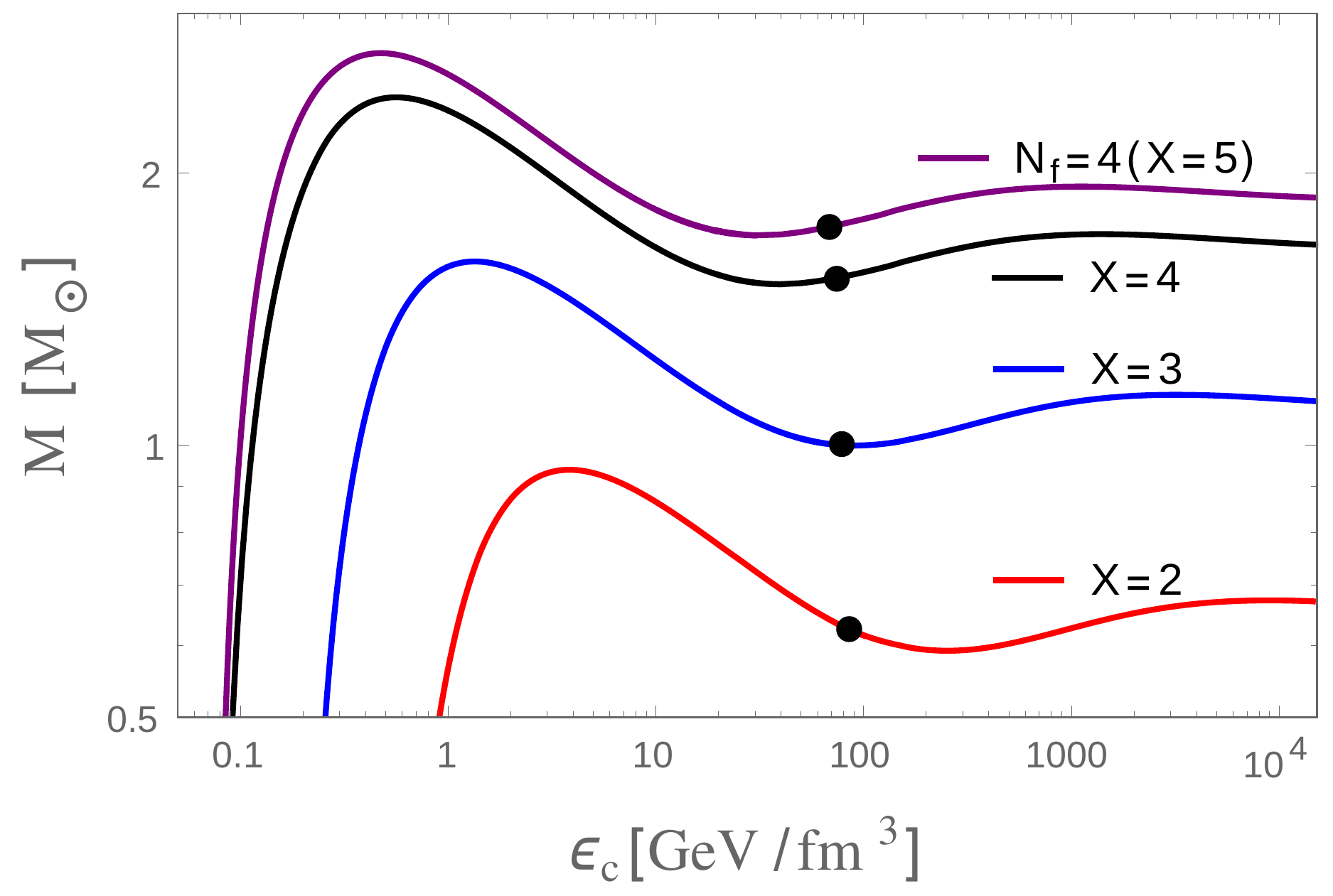}}
     \end{center}
      \caption{\label{fig:gravmass_centraldensity} Total gravitational mass vs central energy density for a system with $N_{f}=2+1+1$ quarks plus leptons. The black dots indicate the appearance of the charm quarks in the system.}
    \end{figure}

\begin{figure}[ht]
      \begin{center}
	\resizebox*{!}{5.5cm}{\includegraphics{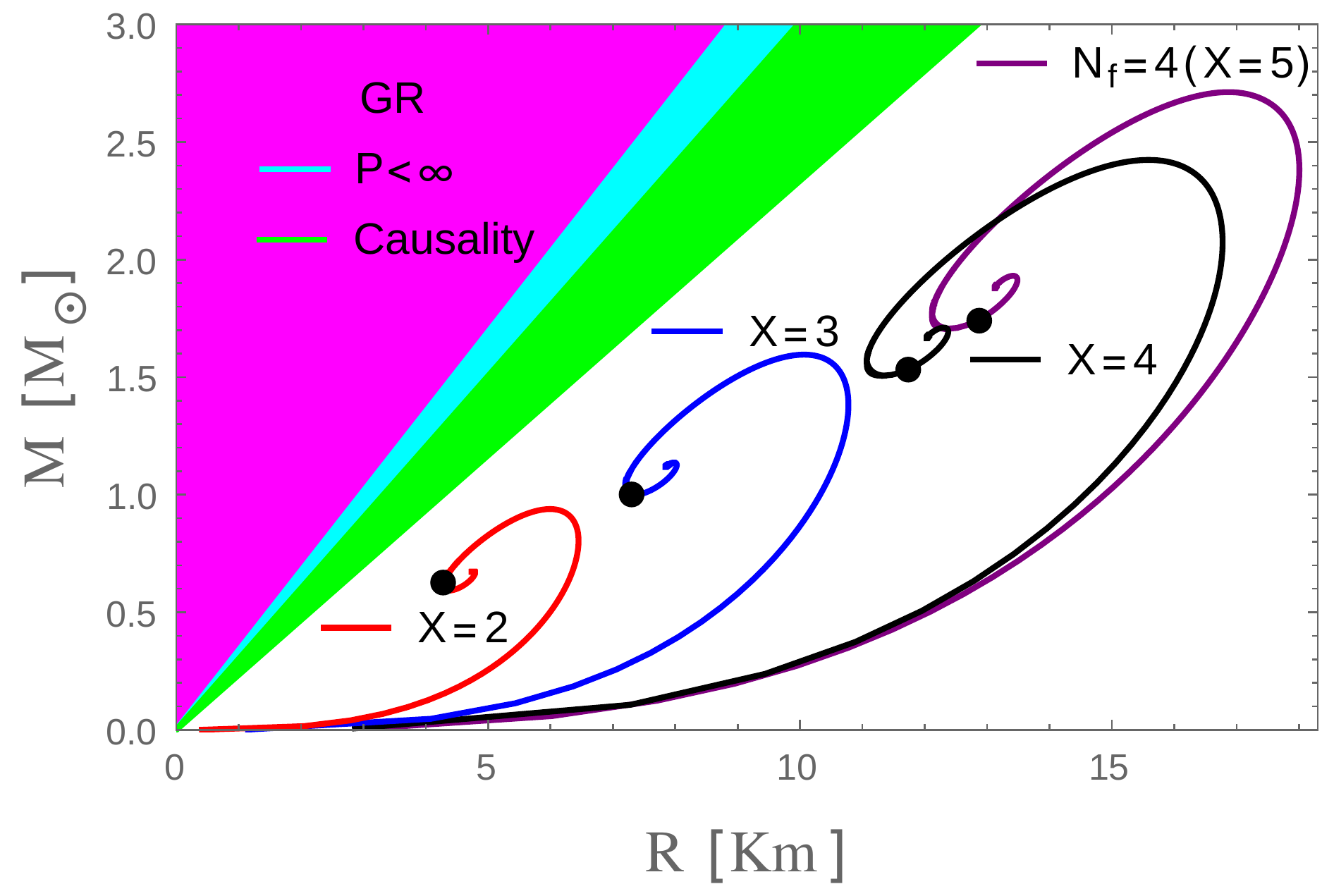}}
     \end{center}
      \caption{\label{fig:mass_radiiNf4} Mass-radius diagram for quark stars made of $N_{f}=2+1+1$ quarks plus electrons and muons. The black dots signal the appearance of charm quarks in the system indicating possible charm star configurations. Stars not satisfying the general-relativistic (GR, magenta region), causality (green region), and finite pressure ($P<\infty$, light blue region) limits are excluded from this diagram \cite{Lattimer:2006xb}. Notice that this exclusion is less restrictive than the one presented in Ref. \cite{Lattimer:2010uk}, based on maximal star masses.}
    \end{figure}

The Tolman-Oppenheimer-Volkov (TOV) equations ensure the relativistic hydrostatic equilibrium of stellar configurations \cite{Glendenning:2000}. However, these configurations must also satisfy the thermodynamic condition $\partial{M}/\partial{\epsilon_{c}}~{\geq}~0$ \cite{Glendenning:2000}. The maximum mass configuration for a given stellar family is identified with the point where $\partial{M}/\partial{\epsilon_{c}}=0$. In Fig. \ref{fig:gravmass_centraldensity}, we show our results for the mass as a function of the central energy density. It can be seen that the necessary condition for thermodynamic stability is satisfied in the two branches, one at relative low and another at much higher energy densities. However, we note that for the case $X=2$, this condition is not satisfied when charm quarks appear, which is indicated by the black dots in Fig. \ref{fig:gravmass_centraldensity}. This is somewhat expected since it is difficult to have heavy quarks present in low-mass strange stars. In Table \ref{tab:table}, we show the values of these observables at the charm threshold. On the other hand, for $X>3$, the thermodynamic condition is satisfied when charm quarks are present, which would correspond to charm stars. In Fig. \ref{fig:mass_radiiNf4}, we show the mass-radius diagram for quark stars made of $N_{f}=2+1+1$ quarks plus electrons and muons for different values of $X$. 

\begin{table}[h!]
  \begin{center}
   \begin{tabular}{c|c|c|c|c} 
     $X$ & $\mu^{\rm th}_{s}[{\rm GeV}]$ & $\epsilon^{\rm th}_{c}[{\rm GeV/fm^{3}}]$ & $M^{\rm th}[\rm M_{\odot}]$ &$R^{\rm th}[{\rm km}]$\\
     \hline
      $2$ & $1.377$ & $85.702$ &$0.625$ & 4.282\\
      $3$ & $1.340$ & $77.988$ &$0.999$ & 7.310\\
      $4$ & $1.290$ & $74.189$ &$1.531$  & 11.72\\
      $5$ & $1.295$ & $69.199$ &$1.745$  & 12.87\\
    \end{tabular}
        \caption{Different values for threshold $\mu_{s}$, $\epsilon_{c}$, gravitational mass $M$ and its associated radii for different values of $X$.}
     \label{tab:table}
  \end{center}
\end{table}

The previous analysis provides a necessary but insufficient condition for stability of star configurations. One must still test for \textit{dynamical} stability under radial pulsations. For that, we use the method of Gondek \textit{et al.} \cite{Gondek:1997fd}, solving a pair of first-order differential equations, one for the relative-displacement variable, $\xi~{\equiv}~\Delta{r}/r$, and another for the Lagrangian perturbation, $\Delta{P}$, with appropriate boundary conditions (see the Appendix). In the first-order radial pulsation formalism, amplitudes oscillate harmonically when the frequencies are such that ${\rm Re(\omega_{n})>0}$ and ${\rm Im(\omega_{n})=0}$, or increase exponentially if\footnote{In this formalism, the maximum mass stellar configuration is characterized by having $\omega_{0}=0$ \cite{Glendenning:2000}.} ${\rm Re(\omega_{n})~{\geq}~0}$ and ${\rm Im(\omega_{n})>0}$. Since the radial oscillation equations represent a Sturm-Liouville problem \cite{Gondek:1997fd}, their eigenvalues (the eigenfrequencies) satisfy the ordering $\omega^{2}_{0}<\omega^{2}_{1}<\omega^{2}_{2}<\cdot\cdot\cdot< \omega^{2}_{n}$. So, if ${\rm Im(\omega_{0})>0}$ [and ${\rm Re(\omega_{0})=0}$] from some value of central energy density $\epsilon_{c}$, then all the higher modes will become complex too, representing the onset of the instability. For convenience, we express our results in the terms of the linear frequency defined by $f_{n}~{\equiv}~\omega_{n}/(2\pi)$, So, if ${\rm Im(f_{0})>0}$ continues finite and increasing for higher densities, all the bare charm quark star configurations become unstable. 

    \begin{figure}[ht]
      \begin{center}
	\resizebox*{!}{5.5cm}{\includegraphics{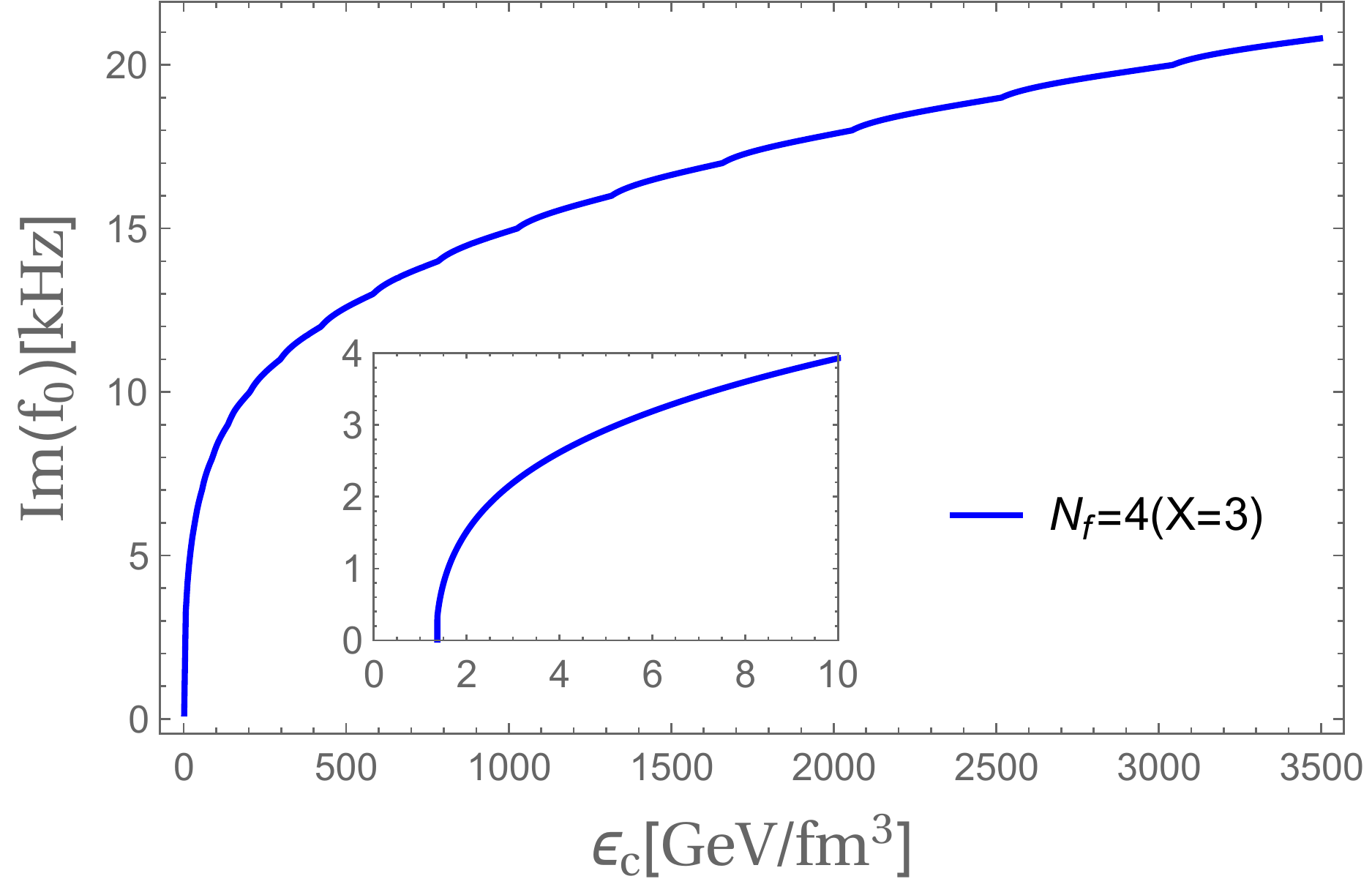}}
     \end{center}
      \caption{\label{fig:ImFreq0X3_density} Imaginary part of the fundamental mode frequency, $f_0$, as a function of the central energy densities obtained after solving the radial pulsation equations for $X=3$.}
    \end{figure}

In Fig. \ref{fig:ImFreq0X3_density}, we show that for densities above the maximum-mass strange star configuration (for $X=3$), the stellar configurations increase their amplitudes even in the region where charm stars are expected, thus making them dynamically unstable. Since the same behavior was obtained for larger values of $X$, one can conclude from a perturbative QCD analysis that charm stars are unstable. One could ask if higher-order perturbative terms could in some way stabilize charm stars. However, a recent N$^{3}$LO weak coupling expansion, which also includes nonperturbative terms, yielded minor modifications to the EoS \cite{Gorda:2018gpy}.

\section{Summary and outlook}
  \label{sec:conclusion}  
  
In this paper we have extended the perturbative QCD $N_{f}=N_{l}+1$ formalism in order to allow for the inclusion of heavy quark flavors in the EoS for cold and dense quark matter and study their effects at low densities by means of the renormalization scale parameter $\bar{\Lambda}$. In particular, we have investigated the effects of charm quarks in the equation of state in the case of $\beta$ equilibrium and electric charge neutrality, where a non-negligible range of the parameter space was discarded in order to go through the charm threshold in agreement with the EoS for light quarks. Then, we have explored the possibility of charm (quark) stars, spanning a range in quark chemical potentials where pQCD is in principle much more reliable. After performing a radial stability analysis, it was concluded these stars would be unstable. 
  
Although charm stars are excluded by our analysis, and also due to causality limits posed by maximum mass constraints from neutron star observations \cite{Lattimer:2010uk,Lattimer:2019eez}, it is possible to have small amounts of charm quark matter in the core of the heaviest observed neutron stars (or, rather, hybrid stars), where a matching between a nuclear and a quark phase could be possible via the Glendenning construction for first-order phase transitions \cite{Han:2019bub}. Recently, a related possibility was investigated under the consideration of strange quark matter contaminated by charm quark impurities (in the sense of condensed matter physics), producing a QCD Kondo effect \cite{Yasui:2013xr,Macias:2019vbl}.  Moreover, a non-negligible amount of charm quarks could contribute to the EoS at the early stage of neutron star mergers, when very high densities are reached \cite{Alford:2017rxf,Most:2018eaw}.
  
Our extended framework is appropriate to study the \textit{heavy} sector of the QCD phase diagram (see Ref. \cite{Maelger:2018vow} for related studies) which could exhibit new features, although it was shown in Refs. \cite{Fischer:2014ata,Burger:2018fvb} that heavy quarks affect negligibly the chiral and deconfinement transitions at finite temperature.   
	
\begin{acknowledgments}
We thank J\"{u}rgen Schaffner-Bielich for very fruitful comments. 
This work was partially supported by CAPES (Finance Code 001), CNPq, FAPERJ, and INCT-FNA (Process No. 464898/2014-5).
\end{acknowledgments}

\section*{Appendix: Radial Pulsation Equations}
	
Assuming static and spherically symmetric charm stars, it is natural to use a Schwarzschild-like line element, having as nontrivial metric functions $e^{\nu(r)}$ and $e^{\lambda(r)}$ for the temporal and radial parts, respectively. After introducing this line element into Einstein's equations and by modeling the star's interior as a perfect fluid, we obtain stellar configurations in hydrostatic equilibrium for a given EoS, governed by the Tolman-Oppenheimer-Volkov (TOV) equations \cite{Glendenning:2000}. 

For the radial pulsation stability analysis, we use the two first-order differential equations of Gondek $\textit{et al.}$ \cite{Gondek:1997fd} for the relative radial displacement $\xi\equiv\Delta{r}/r$ and the Lagrangian perturbation of the pressure $\Delta{P}$, considered independent variables.

Physical smoothness at the star's center requires that the coefficient of the $1/r$ term vanishes for $r\to 0$. So, we  impose that $(\Delta{P})_{\rm center}=-3(\xi\Gamma{P})_{\rm center}$ and normalize the eigenfunctions at the origin to be $\xi(0)=1$. Since $P(r\to R)\to 0$, the Lagrangian perturbation of the pressure at the surface vanishes, i.e., $(\Delta{P})_{\rm surface}=0$. Our code to study stellar charm stars reproduces the pulsation frequencies for the EoSs listed in Ref. \cite{Kokkotas:2000up}.


\end{document}